\documentclass[twocolumn,showpacs,preprintnumbers,amsmath,amssymb,superscriptaddress]{revtex4-2}
\usepackage{graphicx}
\usepackage{dcolumn}
\usepackage{bm}
\usepackage{CJK, CJKnumb}
\begin{document}

\title{Electrical control of crossed Andreev reflection and spin-valley switch in antiferromagnet/superconductor junctions}
\author{Wei-Tao Lu} \email{Email address of Wei-Tao Lu: physlu@163.com.}
\affiliation{School of Physics and Electronic Engineering, Linyi University, 276005 Linyi, China}
\author{Qing-Feng Sun}
\affiliation{International Center for Quantum Materials, School of Physics, Peking University, Beijing, 100871, China}
\affiliation{Collaborative Innovation Center of Quantum Matter, Beijing 100871, China}
\affiliation{Beijing Academy of Quantum Information Sciences, West Bld. $\#$3, No. 10 Xibeiwang East Road, Haidian District, Beijing 100193, China}

\begin{abstract}
We study the subgap transport through the antiferromagnet/superconductor (AF/S) and antiferromagnet/superconductor/antiferromagnet (AF/S/AF) junctions controlled by electric field in a generic buckled honeycomb system, such as silicene, germanene, and stanene. In the present of electric field and antiferromagnetic exchange field, the spin-valley polarized half metallic phase can be achieved in the honeycomb system due to the spin-orbit coupling, which affords an opportunity to generate the pure crossed Andreev reflection (CAR). It is found that the pure CAR can be generated without local Andreev reflection (AR) and elastic cotunneling (EC) over a wide range of electric field. A spin-valley switch effect can be realized between the pure CAR and the pure EC by adjusting the electric field. The properties of AR process and CAR process strongly depend on the spin-valley polarized states. Our results suggest that the device can implement an electrical measurement of the CAR process and spin-valley switch.
\end{abstract}
\maketitle

\section{Introduction}

Andreev reflection (AR) at the interface of a normal metal (N) and superconductor (S) has attracted much attention, where an incident electron in the N is reflected at the interface as a hole and a Cooper pair is injected into the S \cite{Andreev, Beenakker}. The reflected hole retraces the path of the incident electron in the conventional AR, and it is also named Andreev retroreflection. Beenakker predicted that a specular AR could happen at the N/S interface in graphene when an electron in the conduction band is converted into a hole in the valence band \cite{Beenakker, Beenakker2}. In the graphene-based ferromagnet/superconductor (F/S) junction, the ferromagnetic exchange interaction can suppress Andreev retroreflection but enhance the specular AR \cite{Xing}. Sun et al. proposed a four-terminal graphene-superconductor device where the retroreflection and specular reflection could be effectively controlled and separated \cite{Sun}.

The crossed Andreev reflection (CAR) is a nonlocal AR where the conversion from an electron into a hole occurs at two different interfaces of the S. The reverse process of CAR is proposed to be a natural method to generate the entangled electronic state by spatially splitting the Cooper pairs, which has potential application in quantum information and spintronics \cite{Recher, Lesovik, Horodecki, Tan, Herrmann, Burset}. Experimentally, the Cooper pair splitters taking advantage of CAR have been achieved in graphene where a Cooper pair splitting efficiency of $10\%$ is observed \cite{Tan}, and the efficiency is close to $50\%$ in carbon nanotubes \cite{Herrmann}. It is predicted that the splitting efficiency could rise to $100\%$ in the nonlinear regime of N/S/N junction based on carbon nanotubes \cite{Burset}. However, the CAR process often competes with other undesired processes such as the local AR and the elastic cotunneling (EC), making it difficult to distinguish the CAR alone in experiments. Recently, many theoretical \cite{Deutscher, Cayssol, Beenakker3, Benjamin, Linder, Veldhorst, Majidi, Ang, Islam, Soori, Zhang, Gomez, Reinthaler, Sun2, Zhang2, Sun3, Xing2, Wang, Sun4} and experimental \cite{Beckmann, Russo} works have been reported to increase the fraction of the CAR. Cayssol predicted that a pure CAR could be observed in a bipolar graphene transistor due to the unique relativistic band structure of graphene \cite{Cayssol}. However, the complete cancellation of EC and local AR is achieved only at precise biasing to the Dirac point. Veldhorst et al. found that a $100\%$ fraction of the CAR could be achieved in an n-type semiconductor/S/p-type semiconductor, where EC and local AR are fully blocked in the whole band gap region \cite{Veldhorst}. One may create a spin-switch effect between perfect EC and perfect CAR in a superconducting graphene spin valve \cite{Linder}. The N/S/N junction in graphene could act as a Veselago lens which could distinguish the EC signal and CAR signal by moving the focusing point \cite{Gomez}. The topologically protected edge state in topological insulators provides a feasible opportunity to realize the pure CAR \cite{Reinthaler, Sun2, Zhang2, Sun3, Xing2, Wang, Sun4}. The CAR can be measured all-electrically in a quantum spin Hall insulator/S/quantum spin Hall insulator junction \cite{Reinthaler}. A quantized perfect CAR could be obtained due to the tunneling of the chiral Majorana edge states in quantum anomalous Hall insulator/S junctions \cite{Sun3}.

On the other hand, valleytronics has become an important topic of condensed matter physics since the discover of graphene \cite{Beenakker4, Schaibley}. Similar to spintronics, valleytronics aims to manipulate the valley degree of freedom and has great application in quantum information \cite{Schaibley}. Substantial works on valleytronics have been made so far, such as valley-polarized quantum anomalous Hall effect \cite{Ezawa, Yao} and valley transport \cite{Niu, Yokoyama, Lu, Sun5, Lu2}. In graphene like materials, the two degenerate valleys $K$ and $K'$ in the first Brillouin zone are related to each other by time reversal symmetry. The superconducting Cooper pair should be composed of electrons from the two opposite valleys, and so the valley-selective CAR process is expected \cite{Beenakker5, Qi, Yokoyama2, Li, Paul, Chang, Majidi2, Wang2}. The valley polarization of quantum Hall edge states can be detected by a superconducting contact in graphene \cite{Beenakker5}. It is possible to generate a fully spin-valley polarized supercurrent and CAR signal without any contamination from EC or local AR in superconducting silicene \cite{Yokoyama2}. The properties of CAR conductance through the F/S/F hybrid structures are discussed in graphene \cite{Linder, Majidi}, silicene \cite{Li}, and transition metal dichalcogenides \cite{Majidi2}. It is demonstrated that the spin- and valley-switch effects between the pure CAR and pure EC can be controlled by reversing the magnetization direction in one F region \cite{Linder, Li, Majidi2}.

Inspired by the previous literatures, a pure CAR and spin-valley switch which can be controlled electronically are proposed in this work. We study the quantum transport through the AF/S and AF/S/AF junctions in a buckled honeycomb system, such as silicene, germanene, and stanene, which is a monolayer with a buckled honeycomb lattice structure and its band gap could be controlled by electric field. Antiferromagnetic materials have great potential in reducing the device size and power consumption, and AF is expected to replace F as the active spin-dependent element on which spintronic devices are based \cite{Baltz}. In Josephson junctions with AF, the AR could induce Andreev bound states and $0-\pi$ transitions \cite{Bobkova, Andersen}. The dc Josephson current manifests a remarkable atomic-scale dependence on the thickness of AF \cite{Andersen}. Due to the emergence of Specular AR and Andreev retroreflection, the electrical and thermal conductance of AF/S junctions are qualitatively different from those of N(F)/S junctions \cite{Jakobsen}. However, the study of AF/S/AF junctions in the buckled honeycomb system has yet to be explored. In this paper we find that the AF and electric field offer an opportunity to tuning the spin-valley dependent band gap due to the spin-orbit coupling, rendering the system to be spin-valley polarized half metallic. Thus, the proposed junction can be regarded as a half metal/S/half metal junction. It is found that the nonlocal transport can be switched between the pure CAR and pure EC processes by the electric field, the mechanism of which is different from the ones in previous studies by reversing the magnetization direction \cite{Linder, Li, Majidi2}. Furthermore, one may create electrically a spin-valley switch effect between EC and CAR.

The rest of the paper is organized as follows. In Sec. II, we establish the theoretical framework for AF/S and AF/S/AF junctions in the honeycomb system. In Sec. III, we present the main findings for the subgap transport. Finally, we summarize in Sec. IV.

\section{Model and Theoretical Formulation}

\begin{figure}
\includegraphics[width=8.0cm,height=5cm]{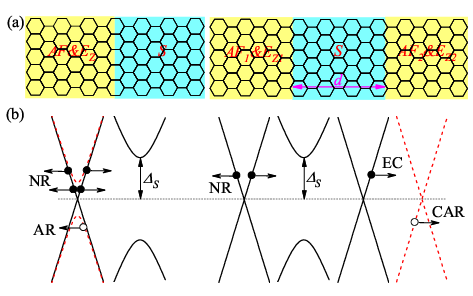}
\caption{ (a) Schematic diagram of the AF/S junction and AF/S/AF junction in the buckled honeycomb system. (b) The energy band in the AF and S regions to explain specular CAR, specular AR, EC, and NR. The black curves in the AF region are the bands for a certain spin at a certain valley, while the red curves are the bands for the opposite spin at the opposite valley.}
\end{figure}

We consider the AF/S and AF/S/AF structures formed on the surface of the buckled honeycomb system, as shown in Fig. 1(a). Via the proximity effect, the superconductivity and the antiferromagnetic exchange field can be produced by depositing the s-wave superconductor \cite{Heersche, Du} and the antiferromagnetic insulators \cite{Tanaka} on the top of the honeycomb sheet, respectively. The width of S region in AF/S/AF junction is $d$. In the present of antiferromagnetic exchange field and electric field, the specular AR and specular CAR processes are schematically displayed by the energy band in Fig. 1(b). The black curves are the bands for a certain spin at a certain valley, while the red curves are the bands for the opposite spin at the opposite valley. An incident electron from the conduction band with a subgap energy $0<E<\Delta_S$ could undergo four possible scattering events. It can either be normally reflected as an electron in the conduction band via normal reflection (NR), or be Andreev reflected as a hole in the valence band with opposite spin and valley via specular AR. It can also be transmitted as an electron via EC or as a hole with opposite spin and valley via specular CAR.

Considering the antiferromagnetic field and superconducting proximity effect, the buckled honeycomb system consisting of two sublattices $A$ and $B$ can be well described by the tight-binding formalism \cite{Yokoyama2, Ezawa2, Yao2}:
\begin{eqnarray}
&&  H = -t \sum_{\langle i,j \rangle, \alpha} c_{i \alpha}^{\dag} c_{j \alpha} + \frac{i \lambda_{SO}}{3\sqrt{3}}
    \sum_{\langle\langle i,j \rangle\rangle, \alpha, \beta} v_{ij} c_{i \alpha}^{\dag} \tau^z_{\alpha\beta} c_{j \beta} \nonumber\\
&&  \lambda_{AF} \sum_{i,\alpha} \xi_i c_{i \alpha}^{\dag} \tau^z_{\alpha\alpha} c_{i \alpha} + l E_z \sum_{i,\alpha} \xi_i c_{i \alpha}^{\dag} c_{i \alpha}  \\
&&  -\mu \sum_{i, \alpha} c_{i \alpha}^{\dag} c_{i \alpha} + \sum_{i, \sigma} (\sigma \Delta_S c_{i \sigma}^{\dag} c_{i -\sigma}^{\dag} + H.c.), \nonumber
\end{eqnarray}
where the first term represents the nearest-neighbor hopping with the transfer energy $t$. $c_{i \alpha}^{\dag}$ ($c_{i \alpha}$) is the electronic creation (annihilation) operator with spin $\alpha$ at site $i$. The second term represents the spin-orbit coupling $\lambda_{SO}$. $\langle i,j \rangle$ and $\langle\langle i,j \rangle\rangle$ run over the nearest-neighbor and the next-nearest-neighbor hopping sites, respectively. $\tau^z_{\alpha\beta}$ is the Pauli matrix of spin and $v_{ij} = +1 (-1)$ if the next-neighboring hopping is anticlockwise (clockwise) with respect to the positive $z$ axis. The third-fifth terms describe the antiferromagnetic exchange field $\lambda_{AF}$, the perpendicular electric field $E_z$, and the electrostatic potential $\mu$, respectively. $\xi_i=\pm 1$ for $i=A, B$ sites. The last term is the superconducting pairing term with superconducting gap $\Delta_S$ induced by the S.

Since we are interested in the AR by bulk states near the Dirac valleys, a simple low-energy Hamiltonian can be derived from the tight-binding model \cite{Yokoyama2, Ezawa2, Yao2}. In the basis of $\psi_k^\dagger=\{(\psi_{k,\sigma}^A)^\dagger, (\psi_{k,\sigma}^B)^\dagger, \psi_{-k,-\sigma}^A, \psi_{-k,-\sigma}^B\}$, the Dirac-Bogoliubov-de Gennes (DBdG) Hamiltonian for given spin $\sigma$ and valley $\eta$ can be written as \cite{Gennes}:
\begin{eqnarray}
&& H_{DBdG} = \left(\begin{array}{cc}  H(k)  &  \sigma \Delta_S  \\  \sigma \Delta_S^{\dag}  &  -H^\dagger(-k)  \end{array}\right)  \nonumber\\
&& =\left(\begin{array}{cc}  H_0 + \sigma \lambda_{AF} \tau_z  &  \sigma \Delta_S  \\  \sigma \Delta_S^{\dag}  &  -(H_0 + \bar{\sigma} \lambda_{AF} \tau_z)  \end{array}\right),
\end{eqnarray}
where $H(k)$ describes electron excitations with certain valley and certain spin, while $-H^\dagger(-k)$ describes hole excitations with opposite valley and opposite spin. The superconducting gap $\Delta_S$ couples the time-reversed electron and hole states from different spins and valleys. $H_0=\hbar v_F (k_x \tau_x - \eta k_y \tau_y) + (\lambda_Z - \eta \sigma \lambda_{SO})\tau_z - \mu$ with Pauli matrices $\tau_{x,y,z}$ and Fermi velocity $v_F$. $\eta=\pm 1$ denotes the valleys and $\sigma=\pm 1$ is the spin index with $\bar{\sigma}=-\sigma$. $\lambda_Z=\ell E_z$ is the staggered sublattice potential between the atoms at $A$ sites and $B$ sites induced by the electric fields $E_z$ due to the buckled structure. $\mu$ is the Fermi level, which is assumed as $\mu_{AF}=0$ in the AF region and $\mu=\mu_S$ in the S region. In order to ensure the validity of the mean-field approximation, the Fermi level $\mu_S$ in the S region should be much larger than the superconducting gap $\Delta_S$. In the S region with $\lambda_{AF}=0$, the system obeys the time reversal symmetry and one may get $TH(k)T^{-1}=H^\dagger(-k)$ with the time reversal operator $T$. In the AF region, the nonzero $\lambda_{AF}$ would break the system's time reversal symmetry. The buckling height is $2\ell=0.46{\AA}$, $0.68{\AA}$, and $0.84{\AA}$, the Fermi velocity is $v_F=5.5$, $4.6$, $4.9$ in units of $10^5m/s$, and the spin-orbit coupling is $\lambda_{SO}=3.9$, $43$, $100meV$ for silicene, germanene, and stanene, respectively \cite{Yao2, Tsai}. Note that we study a large bulk instead of a nanoribbon, and the electron motion is described by the bulk states at the low energy. However, the AR by edge states in nanoribbons would greatly depend on the lattice orientation \cite{Fertig}. In addition, for the proposed junctions, the length scale of AF and S is much larger than the lattice constant which implies that the intervalley scattering is weak at low-energy regions and could be ignored \cite{Castro}. We set $\hbar v_F=1$ for the brevity of notation in the following calculation. The dispersion relation of the AF region can be written as:
\begin{eqnarray}
E_{e(h)}=\pm \sqrt{(\lambda_Z-\eta \sigma \lambda_{SO}+(-)\sigma \lambda_{AF})^2+k_F^2}-(+)\mu. \nonumber\\
\end{eqnarray}

Given the Fermi energy $E$ and transverse momentum $k_y$, the eigenvectors of DBdG Hamiltonian describing electrons and holes in the AF region can be written as:
\begin{eqnarray}
&& \psi_{AF,e}^{\pm} = [\pm k_e e^{\pm i \eta \theta_e}, \gamma_e, 0, 0]^T e^{\pm i k_{ex} x} / N_e, \nonumber\\
&& \psi_{AF,h}^{\pm} = [0, 0, \mp k_h e^{\pm i \eta \theta_h}, \gamma_h]^T e^{\pm i k_{hx} x} / N_h,
\end{eqnarray}
respectively. The related parameters are defined as
\begin{eqnarray}
&& k_{e(h)}=\sqrt{(E+(-)\mu_{AF})^2-\Delta_{e(h)}^2},\nonumber\\
&& k_{e(h)x}=k_{e(h)}\cos\theta_{e(h)},\nonumber\\
&& N_{e(h)}=\sqrt{2(E+(-)\mu_{AF})\gamma_{e(h)}},\\
&& \gamma_{e(h)}=E+(-)\mu_{AF}-(+)\Delta_{e(h)},\nonumber\\
&& \Delta_{e(h)}=\lambda_Z-\eta \sigma \lambda_{SO}+(-)\sigma \lambda_{AF}. \nonumber
\end{eqnarray}
$\Delta_e$ presents an electrically controllable spin and valley dependent band gap, which plays a key role in the AR and CAR processes. The relation between the incident angle $\theta_e$ and the AR angle $\theta_h$ satisfies $k_e\sin\theta_e=k_h\sin\theta_h$ due to the conservation of the transverse momentum. In the S region, the eigenvectors of DBdG Hamiltonian are given by
\begin{eqnarray}
&& \psi_{S,e}^{\pm}=[u_1, \pm u_1 e^{i \eta \phi_e}, u_2, \pm u_2 e^{i \eta \phi_e}]^T e^{\pm(i\mu_S-\kappa)x} / \sqrt{2}, \\
&& \psi_{S,h}^{\pm}=[u_2, \pm u_2 e^{-i \eta \phi_h}, u_1, \pm u_1 e^{-i \eta \phi_h}]^T e^{\pm(i\mu_S+\kappa)x} / \sqrt{2}, \nonumber
\end{eqnarray}
with $u_{1(2)}=\sqrt{1/2\pm\sqrt{E^2-\Delta_S^2}/2E}$ and $\kappa=\sqrt{\Delta_S^2-E^2}$. The transmission angles satisfy $k_e\sin\theta_e=q_{e(h)}\sin\phi_{e(h)}$ with $q_{e(h)}=\sqrt{(\mu_S\pm\sqrt{E^2-\Delta_S^2})^2-\lambda_{SO}^2}$. For $\mu_S \gg k_e$, we may get $\phi_e=\phi_h=0$.

In the AF/S junction, the wave functions can then be written as
\begin{eqnarray}
&& \Psi_{AF}=\psi_{AF,e}^+ + r_e \psi_{AF,e}^- + r_h \psi_{AF,h}^-, \nonumber\\
&& \Psi_S=t_e \psi_{S,e}^+ + t_h \psi_{S,h}^-.
\end{eqnarray}
Matching the wave functions of AF and S regions at the interface: $\Psi_{AF}|_{x=0}=\Psi_S|_{x=0}$, when $\mu_S \gg k_e$ we obtain the AR and NR coefficients as
\begin{eqnarray}
&& r_h=(2\sigma \zeta \gamma_e k_{ex} \sec \alpha) / D,\\
&& r_e=[(\gamma_h k_{e+}-\gamma_e k_{h-}) + i (k_{e+}k_{h-} - \gamma_e \gamma_h) \tan \alpha] / D, \nonumber
\end{eqnarray}
where $D=(\gamma_h k_{e-}+\gamma_e k_{h-}) + i (k_{e-}k_{h-} + \gamma_e \gamma_h) \tan \alpha$, $k_{e\pm}=k_{ex} \pm i \eta k_y$, $k_{h\pm}=k_{hx} \pm i \eta k_y$, $\alpha=\cos^{-1}(E/\Delta_S)\Theta(\Delta_S-E)-i \cosh^{-1}(E/\Delta_S) \Theta(E-\Delta_S)$, and $\zeta=N_h/N_e$. Then one can derive the AR and NR probabilities:
\begin{eqnarray}
&& R_{\eta\sigma}^h = Re (k_{hx})|r_h|^2 / k_{ex} = Re (k_{hx})(2\zeta \gamma_e k_{ex} \sec \alpha)^2 / k_{ex} Q, \nonumber\\
&& R_{\eta\sigma}^e = |r_e|^2   \\
&& = [(\gamma_h k_{e+}-\gamma_e k_{h-})^2 + (k_{e+}k_{h-} - \gamma_e \gamma_h)^2 \tan^2 \alpha] / Q, \nonumber
\end{eqnarray}
with $Q=(\gamma_h k_{e-}+\gamma_e k_{h-})^2 + (k_{e-}k_{h-} + \gamma_e \gamma_h)^2 \tan^2 \alpha$. They satisfy the identity $R_{\eta\sigma}^h+R_{\eta\sigma}^e=1$. The Andreev conductance can be evaluated by the Blonder-Tinkham-Klapwijk formula \cite{Blonder}:
\begin{eqnarray}
G=\sum_{\eta\sigma} G_{\eta\sigma} \int_0^{\pi/2} (1+R_{\eta\sigma}^h-R_{\eta\sigma}^e) \cos \theta_e d \theta_e,
\end{eqnarray}
where $G_{\eta\sigma}=2 e^2 N_{\eta\sigma}(E)/h$ characterizes the ballistic conductance of the AF/S junction, $N_{\eta\sigma}(E)=W\sqrt{(E+\mu_{AF})^2-\Delta_e^2}/2\pi$ denotes the number of transverse modes, and $W$ labels the width of the junction. It is convenient to introduce the normalized conductance $G/G_0$ with $G_0=\sum_{\eta\sigma} G_{\eta\sigma}$.

In another device, i.e., the AF/S/AF junction, we can write the wave functions as:
\begin{eqnarray}
&& \Psi_{AF1} = \psi_{AF1,e}^+ + r_e \psi_{AF1,e}^- + r_h \psi_{AF1,h}^-, \nonumber\\
&& \Psi_S = a \psi_{S,e}^+ + b \psi_{S,e}^- + c \psi_{S,h}^+ + d \psi_{S,h}^-, \\
&& \Psi_{AF2} = t_e \psi_{AF2,e}^+ + t_h \psi_{AF2,h}^+, \nonumber
\end{eqnarray}
where $r_e$, $r_h$, $t_e$, and $t_h$ correspond to the coefficients of NR, AR, EC, and CAR, respectively, which can be determined by matching the boundary condition of wavefunctions at the interface: $\Psi_{AF1}|_{x=0}=\Psi_S|_{x=0}$, $\Psi_S|_{x=d}=\Psi_{AF2}|_{x=d}$. The transmission probability of the EC and CAR processes can be obtained,
\begin{eqnarray}
&& T_{\eta\sigma}^e = Re (k_{e2x})|t_e|^2 / k_{e1x}, \nonumber \\
&& T_{\eta\sigma}^h = Re (k_{h2x})|t_h|^2 / k_{e1x}.
\end{eqnarray}
Because of the conservation of current, we have $R_{\eta\sigma}^h + R_{\eta\sigma}^e + T_{\eta\sigma}^h + T_{\eta\sigma}^e=1$ for the AF/S/AF junction.

\begin{figure}
\includegraphics[width=6.0cm,height=6cm]{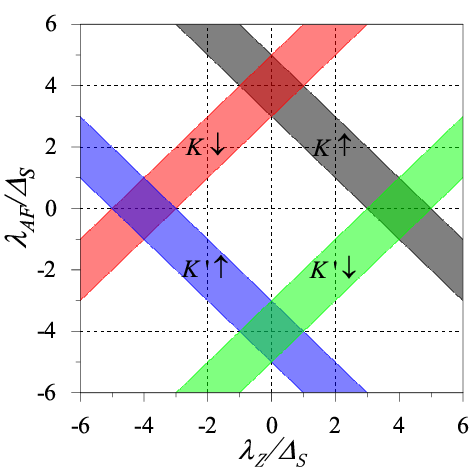}
\caption{ The regions for spin-valley polarized states with a subgap energy $E_{\eta\sigma} \leq \Delta_S$ in the $(\lambda_Z, \lambda_{AF})$ plane with the spin-orbit coupling $\lambda_{SO}=4.0$.}
\end{figure}

Form equations (3) and (5) we can find that the finite $\lambda_{AF}$ and $\lambda_Z$ could destroy the spin and valley degeneracies, leading to the spin-valley polarized states. When $|\Delta_e| \leq |\Delta_S|$, the propagating states can be generated in the AF side. Fig. 2 shows the distribution of spin-valley polarized states with a subgap energy in the ($\lambda_Z$, $\lambda_{AF}$) plane. It can be seen that the four kinds of polarized states are split in different areas, the width of which is $\sqrt{2}\Delta_S$ depending on the Fermi energy. Interestingly, the four areas are crossed in pairs. For example, the area for spin-up states from K valley ($K\uparrow$ states) crosses with that for $K'\downarrow$ states, where $K\uparrow$ states and $K'\downarrow$ states coexist. This feature provides an opportunity for AR. Thus, in the cross area, the AR could appear at the AF/S interface where an incident electron with $K\uparrow$ states (or $K'\downarrow$ states) could be Andreev reflected as a hole with $K'\downarrow$ states (or $K\uparrow$ states). However, in the uncrossed areas, only NR could occur while AR is suppressed. In like manner, AR is also allowed in the cross area of $K\downarrow$ states and $K'\uparrow$ states. Most importantly, one may find that for a given value of $\lambda_{AF}$, the system would undergo four areas for different polarized states in sequence as the electric field $\lambda_Z$ increases. Consequently, for the AF/S/AF junction, the nature of the spin-valley polarized states in the left and the right AF sides could be controlled by the electric fields. Therefore, it is feasible to realized the pure CAR and pure EC which is controlled electrically in the proposed setup. The incident electron with $K\uparrow$ states (or $K\downarrow$ states) can pass through the S region as a hole with $K'\downarrow$ states (or $K'\uparrow$ states) by CAR process, or as an electron with $K\uparrow$ states (or $K\downarrow$ states) by EC process, depending on the electric fields, and so a spin-valley polarized current is formed. The AR, NR, CAR, and EC processes are displayed in Fig. 1(b). There is no propagating bulk state in the white areas of Fig. 2. Interestingly, the quantum spin Hall effect and quantum anomalous Hall effect of nanoribbon system depending on the edge states and Chern numbers could be achieved in the white areas \cite{Ezawa3}. Note that we mainly discuss the AR of bulk states and take no account of edge states since the Fermi energy is in the band. In addition, the AR and CAR processes strongly depend on the Fermi level $\mu_{AF}$. If $\mu_{AF}$ is larger than $\Delta_S$, the retro AR and CAR could occur where both electron and hole lie in the conduction band. The presents work mainly focuses on the specular AR and CAR lead by the interband AR.

\section{Results and discussion}
In the following discussion, we mainly study the local transport in AF/S junction and nonlocal transport in AF/S/AF junction with the subgap energy regime. We concentrate on the specular CAR and specular AR in the junctions as well as the effect of exchange field and electric field. The parameters are set as $\mu_{AF}=0.0$ and $\Delta_S=1.0meV$. Taking silicene for instance, $\lambda_{SO}=4.0\Delta_S$. For convenience, $\Delta_S$ is the unit of $\lambda_Z$, $\lambda_{AF}$, $\lambda_{SO}$, $\mu_S$, and $E$. The unit of width $d$ is the superconducting coherence length $\xi=\hbar v_F/\Delta_S$.

\begin{figure}
\includegraphics[width=8.0cm,height=8.0cm]{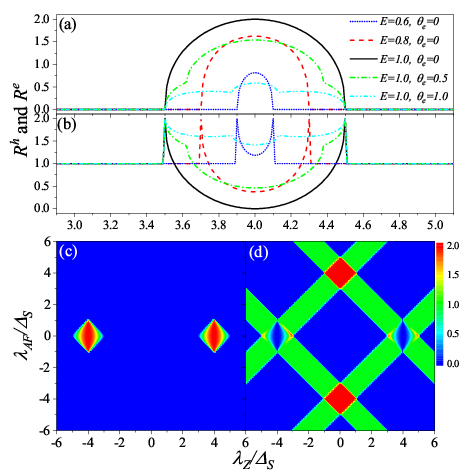}
\caption{ (a) AR probability $R^h$ and (b) NR probability $R^e$ versus $\lambda_Z$ of AF/S junction with $\lambda_{AF}=0.5$. (c) $R^h$ and (d) $R^e$ in the $(\lambda_Z, \lambda_{AF})$ plane at $E=1.0$ for normal incidence.}
\end{figure}

Firstly, we study the AR and NR processes in the AF/S junction. An incoming electron with subgap energy is either normally reflected as an electron or Andreev reflected as a hole. (a) AR probability $R^h$ and (b) NR probability $R^e$ as a function of electric field $\lambda_Z$ are shown in Fig. 3 with $\lambda_{AF}=0.5$ for different values of Fermi energy and incident angle. Here, $R^h$ and $R^e$ are defined as $R^h=\sum_{\eta \sigma} R_{\eta \sigma}^h$ and $R^e=\sum_{\eta \sigma} R_{\eta \sigma}^e$, respectively. One can see that the AR process appears at $3.5 \leq \lambda_Z \leq 4.5$, where $K\uparrow$ states and $K'\downarrow$ states coexist. Specular AR by $K'\downarrow$ (or $K\uparrow$) hole states could occur when $K\uparrow$ (or $K'\downarrow$) electron states are injected to the superconductor interface. In consequence, $R^h=R_{K\uparrow}^h+R_{K'\downarrow}^h$, $R^e=R_{K\uparrow}^e+R_{K'\downarrow}^e$, and $R^h+R^e=2$ in this region. However, only $K\uparrow$ states exist in the region $\lambda_Z<3.5$ while only $K'\downarrow$ states exist in the region $\lambda_Z>4.5$, so that only NR process could happen with $R^e=1$ and AR is suppressed to $R^h=0$, which is clearer in Fig. 4(b). The AR and NR probabilities strongly depend on the Fermi energy $E$ and incident angle $\theta_e$. As $E$ increases, the range for AR probability is broadened which satisfies $\lambda_{SO} - E + \lambda_{AF} \leq \lambda_Z \leq \lambda_{SO} + E - \lambda_{AF}$. When $\lambda_Z=\lambda_{SO}$ and $E=1.0$, AR totally dominates the reflection with $R^h=2$ at normal incidence where the NR is completely forbidden and the electron-hole conversion with unit probability happens. As $\theta_e$ increases, AR probability gradually decreases while NR probability increases. The variations of AR and NR are no longer monotonous for oblique incidence, which can be understood based on the results in Figs. 4(a) and 4(b). In addition, both AR and NR probabilities are symmetric with respect to $\lambda_Z=\lambda_{SO}$ due to the invariance of the spin-valley related band structure.

Figs. 3(c) and 3(d) display the AR probability $R^h$ and NR probability $R^e$ in the ($\lambda_Z$, $\lambda_{AF}$) plane for normal incidence at $E=1.0$, which have an interesting distribution. $R^h$ and $R^e$ are symmetric around $\lambda_Z=0$ and $\lambda_{AF}=0$. Comparing with Fig. 2 one may find that the AR probability only happens in the cross areas of $K\uparrow$/$K'\downarrow$ states and of $K\downarrow$/$K'\uparrow$ states, where the NR probability is strongly suppressed. On the contrary, in the uncrossed areas, only NR process could occur with  unit probability $R^e=1$ completely contributed by one certain spin-valley polarized state while AR process cannot appear. In the cross areas of $K\uparrow$/$K\downarrow$ states and of $K'\uparrow$/$K'\downarrow$ states, the NR probability is $R^e=2$ devoted by two different spin states from one valley. The blue area implies that no AR or NR process could occur, since there is no propagating state in the subgap energy region.

\begin{figure}
\includegraphics[width=8.0cm,height=6.0cm]{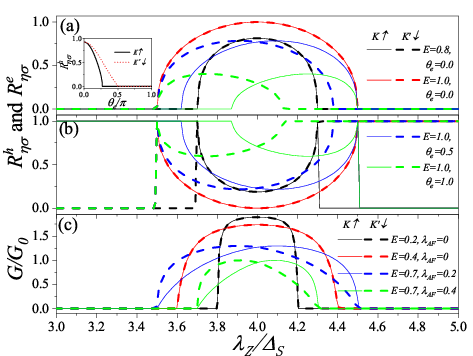}
\caption{ (a) AR probability $R^h_{\eta\sigma}$ and (b) NR probability $R^e_{\eta\sigma}$ versus $\lambda_Z$ for each spin-valley polarized state with $\lambda_{AF}=0.5$. $R_{\eta\sigma}^h$ versus incident angle $\theta_e$ is shown in  the inset of (a) with $\lambda_Z=3.8$ and $E=1$. (c) Andreev conductance $G$ versus $\lambda_Z$ for each spin-valley polarized state in AF/S junction.}
\end{figure}

The total probabilities $R_h$ and $R_e$ by all the polarized states are discussed in Fig.3. In order to study the devotion by each spin-valley polarized state, $R_{\eta\sigma}^h$ and $R_{\eta\sigma}^e$ as a function of $\lambda_Z$ are displayed in Fig. 4 for AF/S junction. In the considered region of $\lambda_Z$, only $K\uparrow$ and $K'\downarrow$ states could devote to AR and NR, while $K\downarrow$ and $K'\uparrow$ states have no contribution. We can see that $R_{K\uparrow}^h=R_{K'\downarrow}^h$ and $R_{K\uparrow}^e=R_{K'\downarrow}^e$ for normal incidence. Note that $R_{K\uparrow}^e=1$ and $R_{K'\downarrow}^e=0$ in the region $\lambda_Z < \lambda_{SO} - E + \lambda_{AF}$ while $R_{K\uparrow}^e=0$ and $R_{K'\downarrow}^e=1$ in the region $\lambda_Z > \lambda_{SO} + E - \lambda_{AF}$ (see Fig. 4(b)). Furthermore, it is important to consider the scattering angles in AR. The incident angle $\theta_e$ and the AR angle $\theta_h$ are related via the relation $k_e\sin\theta_e=k_h\sin\theta_h$. It is implied that there exists a critical angle $\theta_e^c$ above which the AR angle $\theta_h$ exceeds $\pi/2$, the wave vector $k_h$ becomes imaginary, the wavefunction for hole becomes evanescent, and so $R_{\eta\sigma}^h=0$. We can obtain the critical angle $\theta_e^c=\arcsin k_h/k_e$ when $\theta_h=\pi/2$. Distinctly, the critical angle strongly depends on the valley and spin indexes (see the inset of Fig. 4(a)). Therefore, for a nonzero angle $\theta_e$, the AR probabilities as a function of $\lambda_Z$ are different for different valleys and spins. From Fig. 4(a) we can clearly see that $R_{K\uparrow}^h$ and $R_{K'\downarrow}^h$ are no longer the same, and they appear in different regions. Nevertheless, the symmetry between $R_{K\uparrow}^h$ and $R_{K'\downarrow}^h$ around $\lambda_Z=\lambda_{SO}$ still holds. The results suggest that the AF/S junction could work as a valley-spin filter by AR process controlled by incident angle and electric field.

\begin{figure}
\includegraphics[width=8.0cm,height=8.0cm]{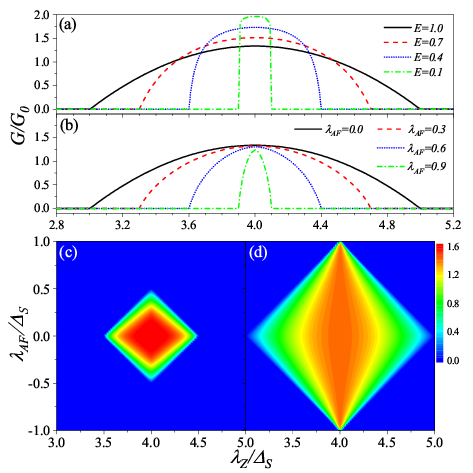}
\caption{ Andreev conductance $G$ versus $\lambda_Z$ of AF/S junction at (a) $\lambda_{AF}=0.0$ and (b) $E=1.0$. Andreev conductance in the $(\lambda_Z, \lambda_{AF})$ plane at (c) $E=0.5$ and (d) $E=1.0$.}
\end{figure}

The dependence of Andreev conductance on the electric field and exchange field is discussed in Fig. 5. We take the cross area of $K\uparrow$/$K'\downarrow$ states as example, where the Andreev conductance is contributed by the $K\uparrow$ states and $K'\downarrow$ states. As shown in Fig. 5(a), the conductance range has an increasing behavior with Fermi energy and satisfies $\lambda_{SO} - E + \lambda_{AF} \leq \lambda_Z \leq \lambda_{SO} + E - \lambda_{AF}$. The conductance is strongly enhanced at low Fermi energy, which approaches to $2G_0$ when $E<0.1$, meaning that the reflection is mainly AR process. Interestingly, the conductance is symmetric with respect to $\lambda_Z=\lambda_{SO}$ and reaches its maximum value at $\lambda_Z=\lambda_{SO}$. This feature can be understood by the behavior of AR and NR probabilities in Figs. 3 and 4. With the increase of the exchange field $\lambda_{AF}$, the conductance range is reduced, as shown in Fig. 5(b). Unlike the ferromagnetic exchange field that enhances the specular Andreev conductance \cite{Xing}, the antiferromagnetic exchange field restrains the specular Andreev conductance. Figs. 5(c) and 5(d) show the properties of the conductance in the ($\lambda_Z$, $\lambda_{AF}$) plane at $E=0.5$ and $1.0$, respectively. Obviously, the conductance range is linearly reduced with $\lambda_Z$ and $\lambda_{AF}$. Moreover, the distribution of conductance is symmetric with respect to $\lambda_Z=\lambda_{SO}$ and $\lambda_{AF}=0.0$. Fig. 4(c) presents the spin-valley polarized Andreev conductance $G$ by $K\uparrow$ states and $K'\downarrow$ states. When $\lambda_{AF}=0$, the conductances by $K\uparrow$ states and $K'\downarrow$ states are the same due to the time-reversal symmetry. The appearance of $\lambda_{AF}$ can break the time-reversal symmetry and destroy the degeneracy of $K\uparrow$ and $K'\downarrow$ states. Accordingly, their conductances are different but symmetric about $\lambda_Z=\lambda_{SO}$. The Andreev conductance has the same properties in the area where $K\downarrow$ states and $K'\uparrow$ states cross.

Next, we further study the nonlocal quantum transport in AF/S/AF junction. As discussed in Fig. 2, owing to the effects of exchange field and electric field, for fixed value of exchange field, taking $\lambda_{AF1}=\lambda_{AF2}=1.0$ for instance, the state with subgap energy in AF1 region can be $K\uparrow$ polarized at $\lambda_{Z1}=3.0$ while other spin-valley polarized states are pushed out of the subgap energy. By adjusting the electric field $\lambda_{Z2}$, the state in AF2 region is $K\uparrow$ polarized at $\lambda_{Z2}=3.0$ and would change to be $K'\downarrow$ polarized when the electric field increases to $\lambda_{Z2}=5.0$. The band structures for spin-valley states are shown in Fig. 1(b). Distinctly, the local AR process in the subgap energy regime is completely suppressed. As a consequence, for incident electron with $K\uparrow$ state propagating to the right, the electrical control of pure CAR by $K'\downarrow$ hole state and pure EC by $K\uparrow$ electron state is expected in AF/S/AF junction.

\begin{figure}
\includegraphics[width=8.0cm,height=7cm]{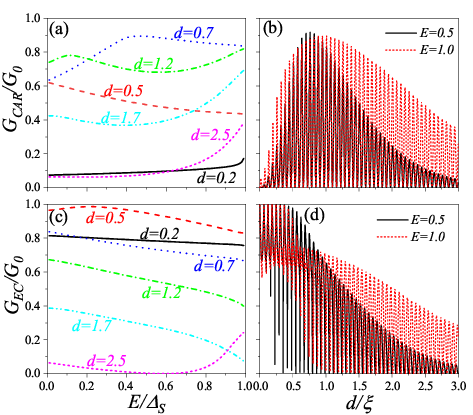}
\caption{ CAR conductance versus (a) $E$ and (b) $d$ at $\lambda_{Z2}=5.0$. EC conductance versus (c) $E$ and (d) $d$ at $\lambda_{Z2}=3.0$. Other parameters are set as $\lambda_{Z1}=3.0$ and $\lambda_{AF1}=\lambda_{AF2}=1.0$. }
\end{figure}

The CAR as a function of (a) Fermi energy $E$ and (b) width $d$ at $\lambda_{Z2}=5.0$ is presented in Fig. 6, where both EC and local AR are forbidden. Thus, we have $R_{\eta\sigma}^e + T_{\eta\sigma}^h = 1$. The nonlocal conductance $G_{CAR}$ for CAR process can be defined as
\begin{eqnarray}
G_{CAR}=\sum_{\eta\sigma} G_{\eta\sigma} \int_0^{\pi/2} T_{\eta\sigma}^h \cos \theta_e d \theta_e.
\end{eqnarray}
It can be seen that $G_{CAR}$ approaches to its maximum when the width $d$ trends to the superconducting coherence length $\xi$. $G_{CAR}$ approaches to zero at $d \ll \xi$ and $d \gg \xi$. The CAR oscillates periodically with $d$. The peaks of $G_{CAR}$ arise from quantum interference, and the conductance peaks appear whenever the subgap superconducting quasiparticle wave vector matches the resonance wave vector $q_e=n\pi/d$. Compared to the silicene F/S/F junction \cite{Li}, the CAR process is present for long width $d$ in the AF/S/AF junction. When $\mu_S \gg E$, $\Delta_S$, and $\lambda_{SO}$, $q_e \approx \mu_S$, so that the oscillation period depends greatly on $\mu_S$. When electric field changes to $\lambda_{Z2}=3.0$, only the EC process is permitted but CAR and AR are forbidden, as shown in Figs. 6(c) and 6(d). In this case, $R_{\eta\sigma}^e + T_{\eta\sigma}^e = 1$. We define the conductance $G_{EC}$ for EC process as
\begin{eqnarray}
G_{EC}=\sum_{\eta\sigma} G_{\eta\sigma} \int_0^{\pi/2} T_{\eta\sigma}^e \cos \theta_e d \theta_e.
\end{eqnarray}
Contrary to $G_{CAR}$, $G_{EC}$ decreases almost monotonically and damply with the width $d$. $G_{EC}$ is close to the maximum value as $d$ approaches zero and to zero when $d \gg \xi$. The magnitude of $G_{CAR}$ and $G_{EC}$ varies strongly with $d$. The oscillation behavior of CAR and EC processes in the AF/S/AF junction is similar to that observed in other systems \cite{Benjamin, Linder, Majidi2}.

\begin{widetext}
\begin{center}
\begin{figure}
\includegraphics[width=13.0cm,height=6.0cm]{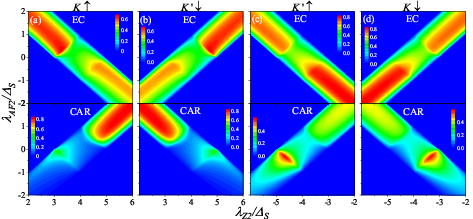}
\caption{ EC conductance and CAR conductance in the $(\lambda_{Z2}, \lambda_{AF2})$ plane for each spin-valley polarized state. $\lambda_{Z1}=3.0$, $5.0$, $-5.0$, and $-3.0$ are for (a), (b), (c), and (d), respectively. Other parameters are set as $\lambda_{AF1}=1.0$, $E=1.0$, and $d=0.7$. }
\end{figure}
\end{center}
\end{widetext}

Fig. 7 exhibits the nonlocal conductances $G_{EC}$ and $G_{CAR}$ in the ($\lambda_{Z2}$, $\lambda_{AF2}$) plane when the electrons with (a) $K\uparrow$ state, (b) $K'\downarrow$ state, (c) $K'\uparrow$ state, and (d) $K\downarrow$ state come from the AF1 lead, which can be realized by adjusting the electric field $\lambda_{Z1}$. Taking $K\uparrow$ electrons for example in Fig. 7(a), compared with Fig. 2 one may find that the EC process exactly happens in the area for $K\uparrow$ state, while the CAR process exactly occurs in the area for $K'\downarrow$ state. In the cross area for $K\uparrow$ and $K'\downarrow$ states, both EC and CAR are found. As a consequence, EC conductance is $K\uparrow$ polarized while CAR conductance is $K'\downarrow$ polarized. The scenarios for other polarized electrons have similar results. The $K'\downarrow$ polarized $G_{EC}$ and $K\uparrow$ polarized $G_{CAR}$ could be achieved in Fig. 7(b) when $K'\downarrow$ electron enters. When $K'\uparrow$ electron enters, the $K'\uparrow$ polarized $G_{EC}$ and $K\downarrow$ polarized $G_{CAR}$ could be achieved (see Fig. 7(c)). On the contrary, the $K\downarrow$ polarized $G_{EC}$ and $K'\uparrow$ polarized $G_{CAR}$ could be obtained when $K\downarrow$ electron enters (see Fig. 7(d)). Furthermore, the EC (or CAR) conductances by $K\uparrow$ and $K'\downarrow$ electrons are symmetric with respect to $\lambda_{Z2}=\lambda_{SO}$, while the EC (or CAR) conductances by $K'\uparrow$ and $K\downarrow$ electrons are symmetric with respect to $\lambda_{Z2}=-\lambda_{SO}$.

\begin{figure}
\includegraphics[width=8.0cm,height=7.0cm]{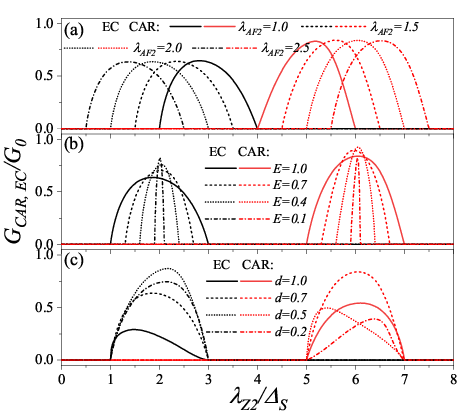}
\caption{ EC conductance and CAR conductance versus $\lambda_{Z2}$. The parameters are set as $\lambda_{Z1}=3.0$, $\lambda_{AF1}=1.0$, $\lambda_{AF2}=2.0$, $E=1.0$, and $d=0.7$, unless otherwise noted in the figure. }
\end{figure}

Finally, we discuss the controllability of CAR and EC processes. Fig. 8 presents the behavior of $G_{EC}$ and $G_{CAR}$ by $K\uparrow$ states as a function of electric field $\lambda_{Z2}$ for different values of (a) $\lambda_{AF2}$, (b) $E$, and (c) $d$ where the AR process is completely forbidden. The conducrances satisfy $G_{CAR}+G_{EC}+G_{NR}=G_0$. Dramatically, the CAR process and EC process occur in separated regions of $\lambda_{Z2}$ with a fixed width $2E$. Importantly, both CAR and EC can be achieved within a wide range of related parameters. The ranges for CAR and EC satisfy $\lambda_{SO}-E+\lambda_{AF2} \leq \lambda_{Z2} \leq \lambda_{SO}+E+\lambda_{AF2}$ and $\lambda_{SO}-E-\lambda_{AF2} \leq \lambda_{Z2} \leq\lambda_{SO}+E-\lambda_{AF2}$, respectively. Therefore, the position and range of pure CAR and pure EC could be realized and modulated by adjusting exchange field $\lambda_{AF2}$ and the Fermi energy $E$ in the present device, as shown in Figs. 8(a) and 8(b). The EC conductance increases as the width $d$ narrows due to the normal tunnel effect of electron (see Fig. 8(c)). $\lambda_{Z1}$ and $\lambda_{AF1}$ have no essential effect to the nonlocal conductance. Whereas, the values of $\lambda_{Z1}$ and $\lambda_{AF1}$ should be located in the regions of $\lambda_{SO}-\lambda_{AF1}-E \leq \lambda_{Z1} \leq \lambda_{SO}-\lambda_{AF1}+E$ and $\lambda_{SO}-\lambda_{Z1}-E \leq \lambda_{AF1} \leq \lambda_{SO}-\lambda_{Z1}+E$, respectively, in order to guarantee that the $K\uparrow$ states in the AF1 side is propagating wave. Furthermore, $G_{EC}$ is a $K\uparrow$ polarized conductance in contrary to the $K'\downarrow$ polarized conductance $G_{CAR}$. Accordingly, the device could work as an electrically controlled spin-valley switch. Note that an electron with $K\uparrow$ state has been mainly considered in order to illustrate the transport, and the cases for other polarized electrons are similar to that for $K\uparrow$ electron.

In order to verify the nonlocal signal in the AF/S/AF junction in experiment, the nonsymmetric voltage could be applied to the junction. Considering that the voltages at the left and right AF leads are $V_L$ and $V_R$, and the superconductor is grounded with the voltage being zero, the current from the left AF lead flowing into the superconductor can be written as \cite{Sun, Sun6}:
\begin{eqnarray}
I_L=\sum_{\eta\sigma} G_{\eta\sigma}\int dE
 [T^{e}_{\eta\sigma}(f_{Le}-f_{Re}) +T^{h}_{\eta\sigma}(f_{Le}-f_{Rh})], \nonumber \\
\end{eqnarray}
when the local AR is zero. Here $f_{Le}(E)=f(E-V_L)$, $f_{Re}(E)=f(E-V_R)$, and $f_{Rh}(E)=f(E+V_R)$ with the Fermi distribution function $f(E)$. At zero temperature, the local and nonlocal differential conductances can be obtained as $G_{local}=dI_L/dV_L =G_{CAR}+G_{EC}$ and $G_{nonlocal}=dI_L/dV_R =G_{CAR}-G_{EC}$, respectively. Obviously, both local and nonlocal conductances are mainly determined by the EC and CAR processes. From the local and nonlocal conductances which can be measured in the experiment, both $G_{EC}$ and $G_{CAR}$ can be obtained.
From Figs. 7 and 8 we can demonstrate that $G_{nonlocal}$ is positive when CAR process dominates and negative when EC process dominates. The nonlocal differential conductance can be abruptly switched from pure EC to pure CAR by changing the electric fields.

\section{Conclusion}

In summary, we studied the local and nonlocal transport through the AF/S and AF/S/AF junctions in a buckled honeycomb system by solving DBdG equation. It is revealed that the AR and NR can be effectively modulated by electric field, due to the spin-orbit coupling and the controllable spin-valley dependent band gap. In the AF/S/AF junction, the spin-valley-selective CAR process is realized. The perfect CAR and perfect EC processes can be switched by adjusting the electric field, which is accompanied by a spin-valley switching effect. Compared with the switch between CAR and EC by reversing the magnetization direction, we proposed an electrical approach which should be more accessible in experiment. The properties of AR process and CAR process by each spin-valley polarized state are also discussed. We expect that these findings are useful to generate the spin-valley entangled states and design the antiferromagnetic device.

\textbf{Acknowledgments:} This work was supported by the NSFC (Grants No. 11974153 and No. 11921005), National Key R and D Program of China (Grant No. 2017YFA0303301), and the Strategic Priority Research Program of Chinese Academy of Sciences (Grant No. XDB28000000).

\end{document}